\documentclass[prd,
onecolumn,
nofootinbib]{revtex4}

\usepackage{graphicx}
\usepackage{longtable,lscape}
\usepackage{array}
\usepackage{bm}
\usepackage{amsmath, amssymb, amsfonts}
\usepackage[usenames]{color}
\usepackage[cp1251]{inputenc}
\usepackage[english]{babel}
\usepackage{color,bm}

\usepackage[a4paper,inner=2cm,outer=1.5cm,top=1.5cm,bottom=2.0cm]{geometry}
\usepackage[colorlinks=true, urlcolor=navyblue, linkcolor=navyblue, citecolor=darkred]{hyperref}
\definecolor{navyblue}{rgb}{0, 0.0, 1.0}
\definecolor{darkred}{rgb}{1.0, 0.0, 0.0}

\newcommand{\vecc}[1]{\mbox{\boldmath $#1$}}
\newcommand{\beq}{\begin{equation}}
\newcommand{\eeq}{\end{equation}}
\newcommand{\beqn}{\begin{eqnarray}}
\newcommand{\eeqn}{\end{eqnarray}}
\newcommand{\be}{\begin{equation}}
\newcommand{\ee}{\end{equation}}
\newcommand{\GEV}{\mbox{GeV}}
\newcommand{\ba}{\begin{eqnarray}}   
\newcommand{\ea}{\end{eqnarray}}

\newcommand{\bc}{\begin{center}} 
\newcommand{\ec}{\end{center}}   

\begin{document}

\title{
Reliability Analysis of the Results of the Known Experiments on Measuring \\ of the Sachs Form
Factor Ratio Using the Rosenbluth Technique. \\ Polarization of the Final Proton in the
$e \vec p \to e \vec p$  Elastic Process
}

\author{M. V. Galynskii} 
\affiliation{Joint Institute for Power and Nuclear Research -- Sosny,
National Academy of Sciences of Belarus, Minsk, 220109 Belarus
}

\begin{abstract}
A criterion for assessing the reliability of measurements of the Sachs form factor
ratio using the Rosenbluth technique is proposed and applied to an analysis
of three known experiments (Andivahis1994, Walker1994, Qattan2005) and
a recent experiment on the CEBAF accelerator upgraded to 12 GeV at JLab
(arXiv:2103.01842 [nucl-ex]). Based on the results of the JLab polarization
experiments on measuring the ratio $\mu_p G_E/G_M$ in the $\vec e  p \to e \vec p$
process, in the kinematics of the SANE Collaboration experiment (2020) on the measurement
of double spin asymmetry in the $\vec e \vec p \to e p$ process numerical
calculations are performed for the $Q^2$ dependence of polarization transferred
to the proton in the $e \vec p \to e \vec p$ process when the initial proton
at rest is partially polarized along the direction of motion of the detected recoil proton.
\end{abstract}

\maketitle

\label{intro}
\section*{INTRODUCTION}
Experiments on studying the proton electric $G_{E}$ and magnetic $G_{M}$ form factors,
the so-called Sachs form factors (SFFs), in the elastic scattering of unpolarized
electrons by protons have been carried out since the mid-1950s \cite{Hofstadter1956}.
All experimental data on the behavior of the SFFs were obtained using the Rosenbluth
technique (RT), which is using the Rosenbluth cross section (in the one-photon
exchange approximation) to the $ep\to ep$ process in the rest frame of the
initial proton \cite{Rosen}
\ba
\label{Ros}
\sigma=
\frac{d\sigma} {d\Omega_e}= \frac{\alpha^2E_2\cos^2(\theta_e/2)}{4E_1^{3}\sin^4(\theta_e/2)}
\frac{1}{1+\tau_p} \left(G_E^{\,2} +\frac{\tau_p}{\varepsilon}\, G_M^{\,2}\right).
\ea
Here, $\tau_p=Q^2/4M^2, Q^2=-q^2=4E_1 E_2\sin^2(\theta_e/2)$ is
the square of the momentum transferred to the proton;
$M$ is the proton mass; $E_1$, $E_2$, and $\theta_e$ are the energies
of the initial and final electrons and the electron scattering
angle, respectively; $\varepsilon$ is the degree of linear polarization
of the virtual photon \cite{Dombey,Rekalo74,AR,GL97} with the variability
domain $0 \leqslant \varepsilon \leqslant 1$,
$\varepsilon=[1+2(1+\tau_p)\tan^2(\theta_e/2)]^{-1}$; and $\alpha=1/137$
is the fine structure constant.

As follows from (\ref{Ros}), at large $Q^2$, the main contribution to the $ep \to ep$ cross
section comes from the term proportional to $G_M^{\,2}$, which already at $Q^2 > 1$ GeV$^2$
makes it rather difficult to extract the contribution of $G_E^{\,2}$. The RT was used
to establish the experimental dependence of SFFs on $Q^2$ described up to $Q^2\approx6$ GeV$^2$
by the dipole approximation \cite{Walker1994,Andivahis1994,Qattan2005,ETG15,Punjabi2015},
and, for their ratio
\be
R \equiv \mu_p G_E/G_M
\label{Rffs}
\ee
the approximate equality $R \approx 1$ holds, where $\mu_p=2.79$
is the proton magnetic moment.

In \cite{Rekalo74}, Akhiezer and Rekalo proposed a method for measuring ratio $R$ based
on polarization transfer from the initial electron to the final proton in the $\vec e  p \to e \vec p$
process. Precision experiments conducted at JLab \cite{Jones00,Gay01,Gay02} using this method
revealed a fast decrease in $R$ with increasing $Q^2$, which indicates SFFs scaling violation.
This decrease in the region $0.4 \, \GEV^2 \leqslant Q^2 \, \leqslant 5.6$ \GEV$^2$
turned out to be linear. Repeated, more accurate measurements of ratio $R$
performed in \cite{Qattan2005,Pun05,Puckett10,Puckett12} in a wide $Q^2$ region up to
8.5 GeV$^2$ using both the Akhiezer–Rekalo method \cite{Rekalo74}
and the RT confirmed the discrepancy of the results.

In \cite{Liyanage2020}, experimental values of $R$ are obtained by the SANE Collaboration
using a third method, which is their extraction from measurements of double spin
asymmetry in the process $\vec e \vec p \to e p$ in the case where the electron-beam
and the proton target are partially polarized. The degree of polarization of the proton
target was $(70 \pm 5)$ \%. The experiment was carried out at two electron-beam
energies of 5.895 and 4.725 GeV and two values of $Q^2$, 2.06 and 5.66 GeV$^2$. The values
of $R$ extracted in \cite{Liyanage2020} agree with the results of the JLab
experiments \cite{Jones00,Gay01,Gay02,Pun05,Puckett10,Puckett12}.

In \cite{JETPL18,JETPL2021}, a fourth method based on polarization
transfer from the initial to the final proton was proposed,
in which $G_E^{\,2}$ and $G_M^{\,2}$ can be extracted from
direct measurements of the cross sections without and
with proton spin flip in the elastic process
\ba
e(p_1)+ \vec{p}\,(q_1,s_1) \to e(p_2)+\vec{p}\,(q_2,s_2)
\label{EPEP}
\ea
when the initial proton (at rest) is completely polarized
along the direction of motion of the final proton
(detected recoil proton). This method also works in
the two-photon exchange (TPE) approximation and
allows squares of modules of generalized SFFs to be
measured in a similar way \cite{JETPL19}.

To resolve the arising contradiction, it was assumed \cite{Melnitchouk2003,Guichon2003}
that the controversy in the experiments could result from the fact that the analysis
ignored higher order radiative corrections, mainly TPE, the effects of which are much
more important in the Rosenbluth technique than in the method \cite{Rekalo74}, because
radiative corrections identically affect the observables of the longitudinal and transverse
polarization of the recoil proton and thus are partially compensated in their ratio.
The TPE contribution to the polarization transfer observables turned out to be small
\cite{Meziane2011}, as the calculations predicted \cite{Melnitchouk2005,Borisyuk2014}.
The idea proposed in \cite{Melnitchouk2003,Guichon2003} stimulated a lot of theoretical
studies of the TPE contribution
\cite{Brodsky2005,Tjon2007,Borisyuk2007,Borisyuk2007-2,Borisyuk2008,Borisyuk2011}
(see reviews \cite{ETG15,Punjabi2015,Carlson2007,Arrington2011,Blunden2017}
and references therein).

The two-photon exchange can be directly extracted by measuring the difference between the cross
sections of the elastic $e^{\pm}p$ scattering processes. These experiments have recently been
conducted at the VEPP-3 storage ring in Novosibirsk \cite{Gramolin2015}, at JLab (the EG5
CLAS experiment \cite{CLAS2015}), and at the DORIS accelerator at DESY (OLYMPUS experiment
\cite{OLYMPUS2017}) with the relevant data available for the region of $Q^2 < 2.1$ GeV$^2$.
Their results have shown that allowances for the TPE may eliminate contradictions at $Q^2$
no larger than 2 GeV$^2$ \cite{Blunden2017}.

The problem of highly accurately measuring the TPE contribution in the extended and so far
largely unexamined region of $Q^2> 2$ GeV$^2$ is supposed to be solved by the CLAS12
experiment \cite{CLASS12} on measuring the $e^{\pm}p$ scattering
cross section ratio using unpolarized $e^{\pm}$ beams from the
CEBAF accelerator upgraded to 12 GeV at JLab. Its results will be decisive
for unambiguously proving or disproving that the TPE is the main source
of discrepancies and for verifying theoretical approaches based
on the consideration of hadron and parton degrees of freedom that may compete
in different parts of the $Q^2$ region under examination.

The polarized positron beam at JLab also provides the unique possibility of
making the first measurement of the polarization
transferred to the proton from the positrons in the
elastic process $e^+ p \to e^+ p$ \cite{CLASS12pos} and a comparison to
the data \cite{Puckett10,Puckett12} on electron scattering may impose
important limits on the hard TPE. The planned experiments
\cite{CLASS12,CLASS12pos} will be an important supplement to the
precision experiment conducted on the upgraded
CEBAF accelerator at JLab \cite{Christy2021} to measure ratio $R$
using the RT at beam energies of 2.2 to 11 GeV and
much larger $Q^2$ of up to 15.75 GeV$^2$ obtained earlier.
The results were analyzed in \cite{Christy2021} using the improved
procedures for calculating total radiative corrections
(RCs) from \cite{Gramolin2016}. Note that the RT-involving experiments
\cite{Walker1994,Andivahis1994} were reanalyzed in \cite{Gramolin2016}, which made it
possible to decrease the values of $R$ measured in those works.

In \cite{Blunden2020}, three well-known experiments
\cite{Walker1994,Andivahis1994,Qattan2005} were reanalyzed in the region
of $Q^2\leqslant 5$ GeV$^2$ using the RT, the improved RC calculation
procedures from \cite{Gramolin2016}, and the TPE contribution calculated
by the authors of \cite{Blunden2020}. Though they were the precision experiments,
and one of them \cite{Qattan2005} even got a special name (Super-Rosenbluth),
the discrepancies between the measurements by the RT and by the polarization
method could be eliminated only for experiment \cite{Andivahis1994} and only in the
region of $Q^2 < 5$ GeV$^2$.

The goal of this work is to try to find out why the results of the experiments
\cite{Walker1994,Andivahis1994,Qattan2005} failed to be reconciled
in \cite{Blunden2020} with the results \cite{Puckett12} and what can come out of
a similar \cite{Blunden2020} reanalysis of the experimental results \cite{Christy2021}.
To this end, a criterion for assessing the reliability of RT measurements of the ratio $R$
is proposed and used to analyze the experimental measurements
\cite{Walker1994,Andivahis1994,Qattan2005,Christy2021}.
Also, based on the results of the JLab polarization experiments on measuring
ratio $R$ in the $\vec e  p \to e \vec p$ process, a numerical analysis
is given to the $Q^2$ dependence of the ratio of the cross sections without and
with proton spin flip and to the polarization asymmetry in the $e \vec p \to e \vec p$
process when the initial (at rest) and final protons are completely polarized and have a
common spin quantization axis coinciding with the direction of motion of the final
proton (detected recoil proton). The longitudinal polarization transferred to
the proton in the case of a partially polarized proton target is calculated in
the kinematics of the SANE Collaboration experiment \cite{Liyanage2020} on measuring double
spin asymmetry in the $\vec e \vec p \to e p$ process.

\label{rest_frame}
\section{
CROSS SECTION FOR THE $e \vec p \to e \vec p$ PROCESS IN THE REST FRAME OF THE INITIAL PROTON
}

Let us consider spin four-vectors $s_{1}$ and $s_{2}$ of the
initial and final protons with four-momenta $q_{1}$ and $q_{2}$
in process (\ref{EPEP}) in an arbitrary frame of reference. The
conditions of orthogonality ($s_{i} q_{i} = 0$) and normalization
($s_{i} ^{2} = - 1$ ) allow their time and space components
$s_i=(s_{i0}, \vecc s_i)$ to be uniquely expressed in terms of
their four-velocities $v_i=q_i/M$ ($i=1, 2$)
\ba
s_i=(s_{i0}, \vecc s_i), \; s_{i0}=\vecc v_i\, \vecc c_i, \;
\vecc s_i =\vecc c_i + \frac{(\vecc c_i \vecc v_i)\,\vecc v_i}{1+v_{i0}}\;,
\label{spinv}
\ea
where the unit three-vectors $\vecc c_i$ ($\vecc c_i^{2}=1$) are the spin
projection axes (spin quantization axes).

In the laboratory reference frame (LRF), where $q_1=(M,\vecc 0)$ and $q_2=(q_{20}, \vecc q_2)$,
we choose spin projection axes $\vecc c_{1}$ and $\vecc c_{2}$ such that they
coincide with the direction of motion of the final proton
\ba
\vecc c = \vecc c_{1} =\vecc c_{2}=\vecc n_2=  \vecc {q_2}/|\vecc q_2|\,.
\label{LSO}
\ea
Then spin four-vectors of the initial ($s_{1}$) and final
protons ($s_{2}$) in the LF take the form
\ba
\label{DSB_LSO1}
s_1=(0,\vecc n_2 )\,, \; s_2= (|\vecc v_2|, v_{20}\, \vecc {n_2})\,,
\,\vecc n_2=  \vecc {q_2}/|\vecc q_2|\,.
\ea

The method \cite{JETPL18} is based on the expression for the
differential cross section of process (\ref{EPEP}) in the LRF when
the initial and final protons are polarized and have a
common spin projection axis $\vecc c$ (\ref{LSO})
\ba
\label{RosPol}
\frac{d\sigma_{\delta_1, \delta_2}} {d\Omega_e}&=&
\omega_{+} \sigma^{\uparrow\uparrow}+\omega_{-}\sigma^{\downarrow\uparrow}\,,\\
\label{RosPol2}
\sigma^{\uparrow\uparrow}&=&\sigma_M \, G^2_E ,\;\;
\sigma^{\downarrow\uparrow}=\sigma_M \frac{\tau_p}{\varepsilon} \, G^2_M\,,\\
\sigma_M&=& \frac{\alpha^2E_2\cos^2(\theta_e/2)}
{4E_1^{\,3}\sin^4(\theta_e/2)} \frac{1}{1+\tau_p}\,.
\ea
Here, $\omega_{\pm}$ are the polarization factors
\ba
\omega_{+}=(1 + \delta_1 \delta_2)/2, \, \,\omega_{-}=(1 -\delta_1 \delta_2)/2\,,
\label{omegi}
\ea
where $\delta_{1,2}$ are the doubled projections of the initial and
final proton spins on the common axis of spin projections $\vecc c$ (\ref{LSO}).
Note that formula (\ref{RosPol}) is valid at
$-1\leqslant \delta_{1,2}\leqslant 1$.

The corresponding experiment on measuring squares of SFFs in the processes
without and with proton spin flip can be implemented as follows. The initial
proton at rest should be completely polarized along the direction of motion
of the final proton (detected recoil proton). Measuring the $Q^{\,2}$ dependence of the
differential cross sections $\sigma^{\uparrow\uparrow}$ and
$\sigma^{\downarrow\uparrow}$ (\ref{RosPol2}), one can also
get information about the $Q^{\,2}$ dependence of $G_E^{\,2}$ and $G^{\,2}_M$
and thus measure them.

Note that formula (\ref{RosPol}), like (\ref{Ros}), can be decomposed
into a sum of two terms involving only $G_E^{\,2}$ and $G^{\,2}_M$.
Averaging and summing (\ref{RosPol}) over polarizations of the
initial and final protons, we obtain a different representation
for cross section (\ref{Ros}), designated as $\sigma_R$ \cite{JETPL18}
\ba
\label{Ross}
\sigma_R =\sigma^{\uparrow\uparrow} + \sigma^{\downarrow\uparrow}.
\ea
Consequently, the physical meaning of the decomposition
of Rosenbluth formula (\ref{Ros}) into a sum of two
terms involving only $G_E^{\,2}$ and $G^{\,2}_M$ is that it is a sum of
cross sections without and with proton spin flip in the
case where the initial proton at rest is completely
polarized along the direction of motion of the final proton.

Note that, in the literature, including textbooks on particle physics, it is often
stated that the use of SFFs is merely convenient to make the Rosenbluth formula
simple and compact. Since these formal considerations
about their advantages also occur in the known
monographs \cite{AB,BLP} written many years ago, they
have not been questioned and have been reproduced in
the literature until now \cite{Paket2015}.

Cross section (\ref{RosPol}) can be written as
\ba
&& d\sigma_{\delta_1, \delta_2}/ d\Omega_e
\label{sigma_d1_d2}
=(1+\delta_2 \delta_f) (\sigma^{\uparrow\uparrow}+\sigma^{\downarrow\uparrow}),\\
\label{delta_f}
&& \delta_f=\delta_1 (R_{\sigma}-1)/(R_{\sigma}+1), \\
&& R_{\sigma}=\sigma^{\uparrow\uparrow}/\sigma^{\downarrow\uparrow},
\label{R_sig}
\ea
where $\delta_f$ is the degree of longitudinal polarization of
the final proton. In the case of a completely polarized
initial proton ($\delta_1=1$), $\delta_f$ coincides with the ordinary
definition of polarization asymmetry
\ba
\label{Asim}
A= (R_{\sigma}-1)/(R_{\sigma}+1)\,.
\ea

As follows from (\ref{RosPol2}), the ratio of the cross sections
without and with proton spin flip $R_{\sigma}$
(\ref{R_sig}) can be expressed in terms of the experimentally measured
quantity $R = \mu_p\, G_E/G_M$
\ba
R_{\sigma}=\frac{\sigma^{\uparrow\uparrow}}{\sigma^{\downarrow\uparrow}}
=\frac{\varepsilon}{\tau_p}\,\frac{G^{\,2}_E}{G^{\,2}_M}=
\frac{\varepsilon}{\tau_p}\frac {R^2}{ \mu_p^2}.
\label{rat2}
\ea

The relative contribution $\Delta_{\sigma}$ of the term $\sigma^{\uparrow\uparrow}$
(\ref{RosPol2}) involving $G^{\,2}_E$ to the cross section $\sigma_R$ (\ref{Ross}) has the form
\ba
\Delta_{\sigma}=\frac{\sigma^{\uparrow\uparrow}}
{\sigma^{\uparrow\uparrow}+ \sigma^{\downarrow\uparrow}}=
\frac {R_{\sigma}}{(1+R_{\sigma})}.
\label{Delta_si}
\ea

We recast formula (\ref{delta_f}) for the degree of polarization
of the final proton in the standard notation,
replacing $\delta_f$ with $P_r$ and $\delta_1$ with $P_t$
\ba
P_{r}= P_t (R_{\sigma}-1)/(R_{\sigma}+1).
\label{Asim_pt}
\ea

With constraint inversion in (\ref{Asim_pt}), we have the
expression for $R^2$ as a function of $P_r/P_t$
\ba
R^2=\mu_p^2\, \frac{\tau_p}{\varepsilon}\,\frac{1+R_p}{1-R_p}, \; \;
R_p=\frac{P_r}{P_t}\,,
\label{Rp}
\ea
which can be used for extracting $R$ in the method of
polarization transfer from the initial to the final proton
\cite{JETPL18,JETPL2021}.

For numerical calculations of the $Q^2$ dependence
of the polarization asymmetry $A$ (\ref{Asim}), cross section
ratio $R_{\sigma}$ (\ref{rat2}), relative contribution $\Delta_{\sigma}$ (\ref{Delta_si}),
and polarization transferred to the proton $P_r$ (\ref{Asim_pt}) in the case of
the dipole dependence ($R=R_d$ ) or its violation
($R=R_j$ ), we will use the parametrization
\ba
\label{Rd}
&&~~~~~~~~~~~~~~~~~~~~~~~~~~R_d=1 ,    \\
&&R_j = \frac{1}{1+0.1430\,Q^2-0.0086\,Q^4+0.0072\,Q^6}.~~~~
\label{Rdj}
\ea

The expression for $R_{j}$ is borrowed from \cite{Qattan2015}, and
the Kelly parametrization \cite{Kelly2004} can be used instead.

\subsection{
RESULTS OF NUMERICAL CALCULATIONS AND THEIR DISCUSSION
}

\begin{figure*}[h!tpb]
\centering
\includegraphics[scale=0.35]{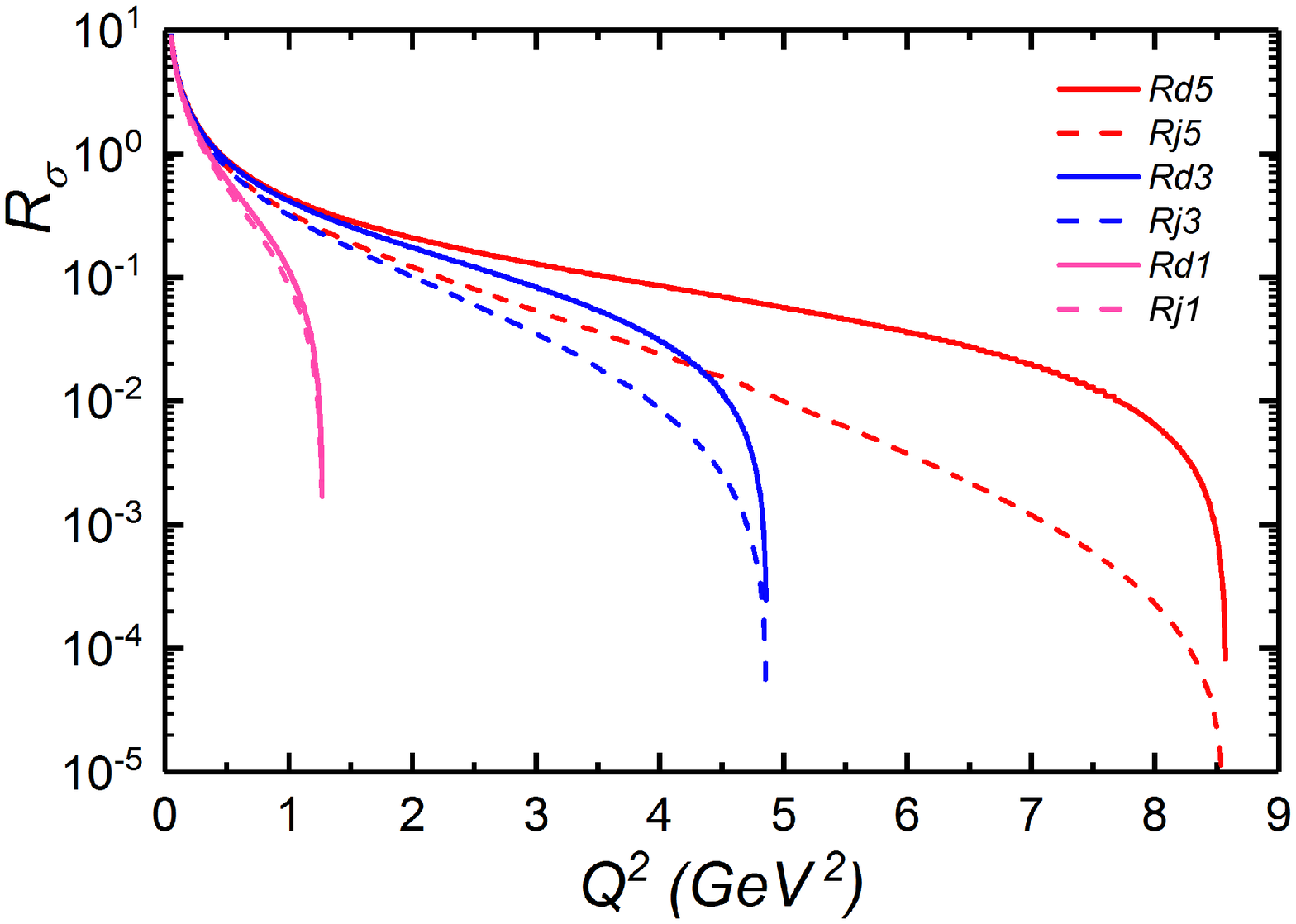}
\includegraphics[scale=0.35]{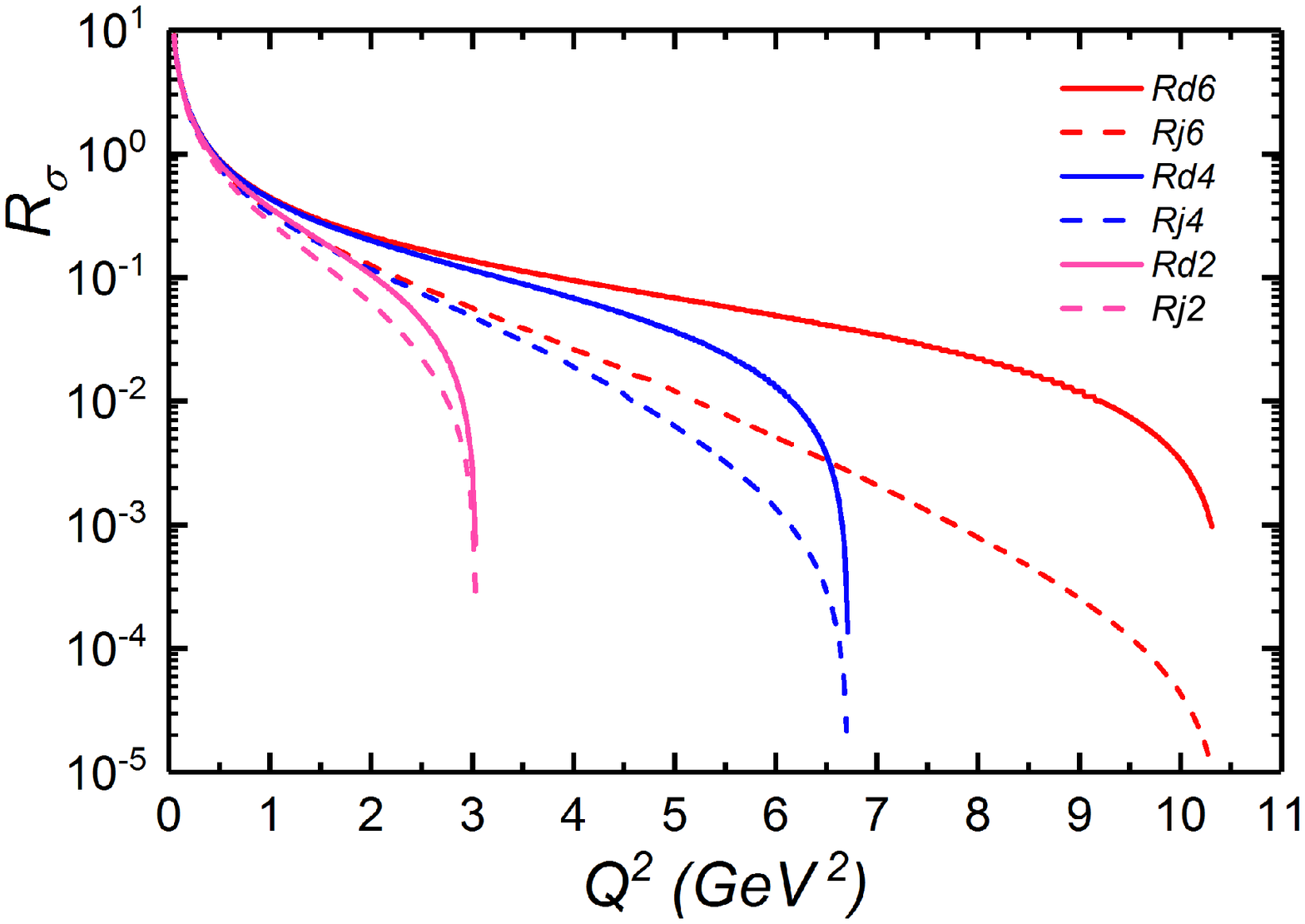}
\vspace{-3mm}
\caption{
Dependence of cross sections ratio $R_{\sigma}$ (\ref{rat2}) on
$Q^2$ (GeV$^2$) for energies $E_1=1, 2, ..., 6$ GeV.
Lines $Rd1$, $ Rd2$,..., $ Rd6$ and $Rj1$, $Rj2$, ..., $Rj6$ correspond
to ratios $R=R_d$ (\ref{Rd}) and $R=R_j$ (\ref{Rdj}).
}
\label{Ratfig}
\end{figure*}

To clarify the general laws, the $Q^2$ dependence of
cross sections ratio $R_{\sigma}$ (\ref{rat2}) was numerically calculated
for the electron-beam energies $E_1=1, 2, ..., 6$ GeV. The
results are plotted in Fig. \ref{Ratfig} for $R=R_d$ (\ref{Rd})
(lines $Rd1$, $ Rd2$,..., $ Rd6$) and $R = R_j$ (\ref{Rdj})
(lines $Rj1$, $Rj2$, ..., $Rj6$). 

It follows from Fig. \ref{Ratfig} that ratios of the cross sections
without and with proton spin flip $R_{\sigma}$ (\ref{rat2})
decrease with increasing $Q^2$ for all electron-beam energies.
However, the decrease at $R=R_j$ is faster than at $R=R_d$
because of the denominator in the expression
for $R_J$ (\ref{Rdj}). Note also that, at low electron-beam energies,
the difference in the behavior of ratio $R_{\sigma}$ (\ref{rat2}) at $R=R_d$
and $R=R_j$ is insignificant.

It is clearly seen in Fig. \ref{Ratfig} that the dependence of $R_{\sigma}$
on $Q^2$ for each electron-beam energy has a sharp
boundary at $Q^2_{max}$, which is the maximum possible
value of $Q^2$ corresponding to the backward (180$^{\circ}$) electron
scattering. Values $Q^2_{max}$ for the beam energies $E_1=1, 2, ... 6$
GeV are presented in Table \ref{Table1}, from which
it follows that $Q^2_{max}$ is no larger than 10.45 GeV$^2$ for all
energies considered.

\vspace{-2mm}
\begin{table}[h!]
\centering
\caption{ 
Values $Q^2_{max}$ determining spectrum boundaries of the $R_{\sigma}$
dependence on $Q^2$ and values $(Q_0)_{\{d,j\}}^2$ at which
$\sigma^{\uparrow\uparrow}=\sigma^{\downarrow\uparrow}$;
polarization asymmetry $A$ (\ref{Asim}) is zero in this case
}
\label{Table1}
\tabcolsep=1.5mm
\vspace{1.0mm}
\footnotesize
\begin{tabular}{|  l | c | c |  c |  c |  c |  c |  c |}
\hline
$ E_1$  (\rm{GeV}) & 1.0 & 2.0 &   3.0 &  4.0 & 5.0 & 6.0  \\
\hline
$ Q^2_{max}$ (GeV$^2$) & 1.277 & 3.040 & 4.868 & 6.718 & 8.578 & 10.443  \\
\hline
$ (Q_0^2)_d$ (GeV$^2$) & 0.358 & 0.424 & 0.435 & 0.446  & 0.446  & 0.446 \\
\hline
$ (Q_0^2)_j$ (GeV$^2$) & 0.336 & 0.380 & 0.391 & 0.402 & 0.402 & 0.402  \\
\hline
\end{tabular}
\end{table}

Table \ref{Table1} also presents the values $(Q_0^2)_{\{d,j\}}$ corresponding
to the equality of the cross sections without and with proton spin flip
$\sigma^{\uparrow\uparrow}=\sigma^{\downarrow\uparrow}$. In this case, their
ratio is $R_{\sigma}=1$ and the polarization asymmetry is zero.
In the case of the dipole dependence, we have $(Q_0^2)_d \approx M^2/2$,
where $M$ is the proton mass. If the dipole dependence is violated, we have
$(Q_0^2)_j \approx 0.40$ GeV$^2$; i.e., equality of the cross sections
$\sigma^{\uparrow\uparrow}$ and $\sigma^{\downarrow\uparrow}$ begins at approximately
the same point where ratio $R$ begins linearly decreasing, thus, the points where
$Q^2=Q_0^2$ are, in a sense, distinguished.

The calculations depicted in Fig. \ref{Ratfig} make it possible to understand
why measurements of ratio $R$ using the RT are faced with difficulties at
large $Q^2$. They should be conducted in the kinematics in which relative contribution
$\Delta_{\sigma}$ (\ref{Delta_si})) of the term $\sigma^{\uparrow\uparrow}$
to cross section $\sigma_R$ (\ref{Ross}) is higher than Rosenbluth cross section measurement
accuracy $\Delta_0$ in this experiment
\ba
\Delta_{\sigma} > \Delta_0.
\label{treb1}
\ea

When $\Delta_0 \ll 1$, inequality (\ref{treb1}) is reduced to
\ba
R_{\sigma} > \Delta_0.
\label{treb3}
\ea

The requirements imposed by inequalities (\ref{treb1}) and (\ref{treb3})
can be considered the necessary conditions for performing reliable measurements.
In the analysis of the experimental results, they can be used as a reliability
assessment criterion for measurements.

The accuracy of Rosenbluth cross section measurements $\Delta_0$ appearing in (\ref{treb1})
and (\ref{treb3}) is determined in the general case by statistical, systematic,
and normalization uncertainties. Below in the reliability analysis of the experimental
measurements \cite{Andivahis1994}, it will be established on the basis
of the results \cite{Blunden2020} that $\Delta_0$ is determined by the normalization
uncertainty. Note that the limits on the kinematics of the experiment conducted using
RT was not considered in the literature, including
\cite{Egle_2018,Egle_2016,Gramolin2016,Blunden2020}. Nevertheless, it seems
to be an important issue that deserves attention.

Tracing lines $Rd1$, $Rd2$,..., $Rd6$ in Fig. \ref{Ratfig}, we make up Table
\ref{Table2} of values $R_{\sigma}$ (\ref{rat2}) for $E_1=1, 2,..., 6$ GeV
and $Q^2=1, 2, ..., 9$ GeV$^2$. In this table, rows (columns) correspond
to the same initial electron-beam energy $E_1$ (square of momentum transferred
to proton $Q^2$).

\vspace{-5mm}
\begin{table*} [h!tpb]
\caption{ 
Values $R_{\sigma}$ (\ref{rat2}) at $R=R_d$ (\ref{Rd}) for electron-beam
energies $E_1=1, 2, ..., 6$ GeV and $Q^2=1,2, ..., 9$ GeV$^2$ }
\label{Table2}
\tabcolsep=1.25mm
\vspace{1.0mm}
\centering
\footnotesize
\begin{tabular}
{| c| c|  c| c|  c|  c|  c|  c|  c|  c|}
 \hline
$E_1$ $\backslash Q^2$
& 1.0 & 2.0 & 3.0 & 4.0 & 5.0 & 6.0 & 7.0 & 8.0 & 9.0  \\
\hline
 6  & 0.444 & 0.215 & 0.136 & 0.095 &
 0.068 & 0.049 &  0.034 &  0.022 &  {0.012} \\
\hline
 5  & 0.440 & 0.209 & 0.129 & 0.086 & 0.057 & {0.036} &  0.020 & 0.006 & \\
\hline
 4  & 0.432 & 0.199 & 0.115 & 0.068 &  {0.037} &  {0.013} & & &  \\
\hline
 3  & 0.415  & 0.175 & 0.084 &  {0.031} &  & & & & \\
\hline
 2  & 0.365 & 0.105 & & & & & & & \\
\hline
 1  & 0.114 & & & & & & & & \\
\hline
\end{tabular}
\end{table*}

For all Table \ref{Table2} cells, except the one with $R_{\sigma}=0.006$,
the relation $R_{\sigma} \geqslant 0.020$ holds at $Q^2=7.0$ and 8.0 GeV$^2$.
Applying criterion (\ref{treb3}), we come to the
conclusion that, at $Q^2=7.0$ GeV$^2$, measurements
with the use of the RT should be performed with an
accuracy no worse than 1.9 \%, and at $Q^2=8.0$ GeV$^2$
the accuracy is required to be $0.3 \div 0.5$ \%. Thus, the difficulties
faced in the experimental measurements of
ratio $R$ at large $Q^2$ using the RT are the decrease in the
relative contribution of the term $\sigma^{\uparrow\uparrow}$ to the Rosenbluth
cross section (\ref{Ross}) and the necessity of increasing the
accuracy of its measurement. Note that, in earlier
experiments, measurements of the Rosenbluth cross
sections with an accuracy higher than 2\% was an
unsolvable problem for many reasons \cite{Bernauer2014}.

\section{Reliability analysis of the experimental results
\cite{Andivahis1994,Qattan2005,Walker1994,Christy2021}}

To analyze reliability of measurements of ratio $R$ in experiments
\cite{Andivahis1994,Qattan2005,Walker1994,Christy2021},
numerical calculations of relative contributions $\Delta_{\sigma}$ (\ref{Delta_si})
were performed for all electron-beam energies $E_1$ and squares of momenta transferred
to the proton $Q^2$ at which measurements were
carried out in \cite{Andivahis1994,Qattan2005,Walker1994,Christy2021}.
The results are presented in Tables \ref{Table3}, \ref{Qattan}, \ref{Walker} and
\ref{Gramolin} respectively. Values $E_1$ (GeV) are given in the first columns,
and $Q^2$ (GeV$^2$) are in the upper rows of the tables. Empty cells
in the tables indicate that measurements were not performed at their corresponding values.

{\bf Reliability analysis of measurements in the experiment \cite{Andivahis1994}}.
For this analysis, we refer to Fig. 15 (b) from \cite{Blunden2020}
depicting results of a reanalysis of the experiments
\cite{Qattan2005,Andivahis1994,Walker1994}, which was performed using the improved procedures
from \cite{Gramolin2016} for the calculation of RCs and the added TPE contribution
calculated in \cite{Blunden2020}. For the reader’s convenience, Figs. 15 (a) and 15 (b)
from \cite{Blunden2020} are given below in Figs. \ref{Blunden}(a), \ref{Blunden}(b),
respectively. It follows from Fig. \ref{Blunden}(b) that measurements
at $Q^2 < 5.0$ GeV$^2$ in \cite{Andivahis1994} with the added TPE contribution
(Andivahis + TPE) agree well with the results \cite{Puckett12}, while at
$Q^2 = 5.0$ GeV$^2$ even the allowance for the TPE fails to eliminate
discrepancies. That is why the measurement corresponding to the bottom cell
of the column for $Q^2 = 5.0$ GeV$^2$ in Table \ref{Table3} is taken
to be unreliable, i.e., insufficiently accurate.

\begin{figure*} [h!tpb]
\centering
\includegraphics[scale=0.25,natwidth=610,natheight=610]{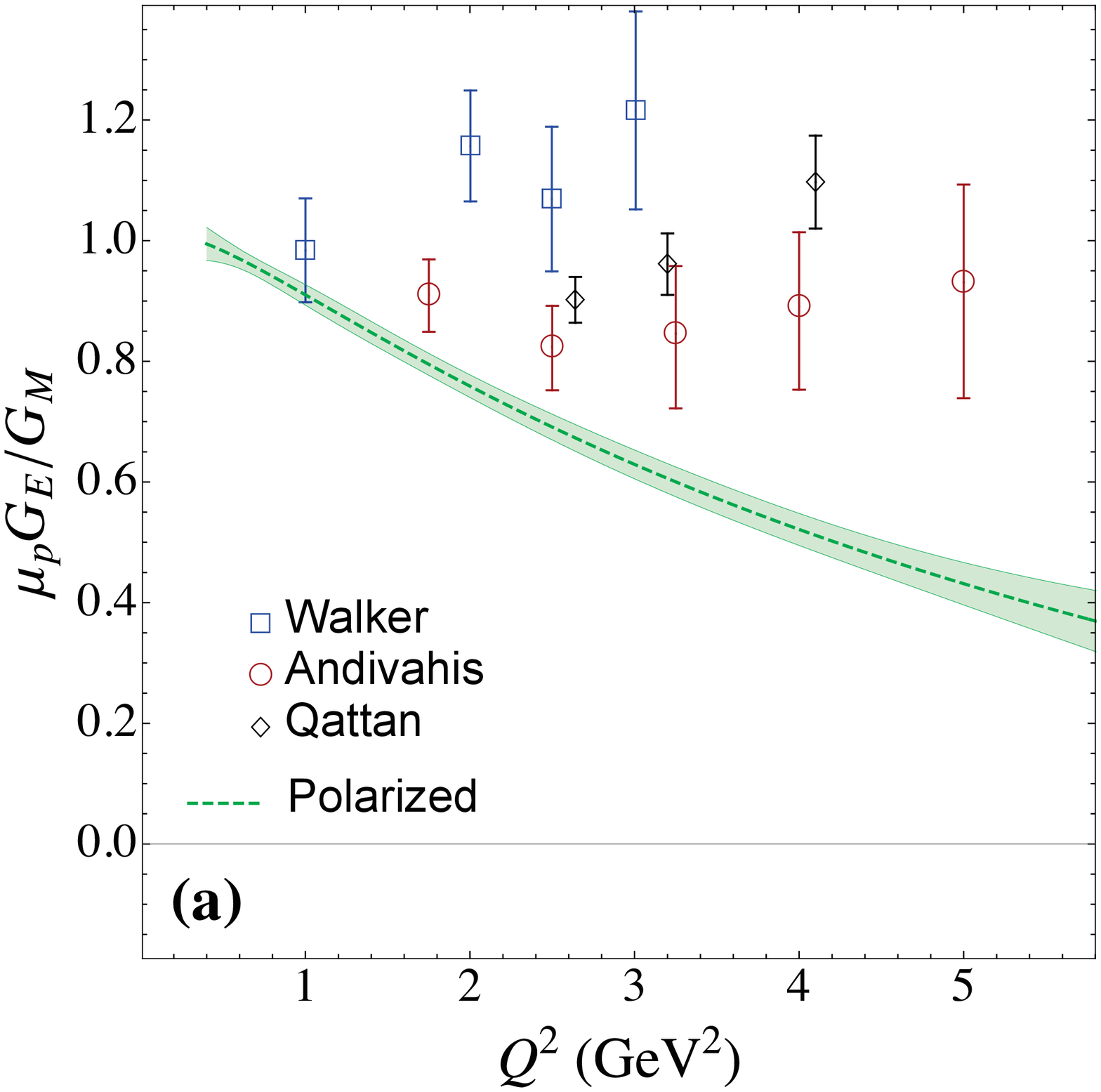}
\includegraphics[scale=0.25,natwidth=610,natheight=610]{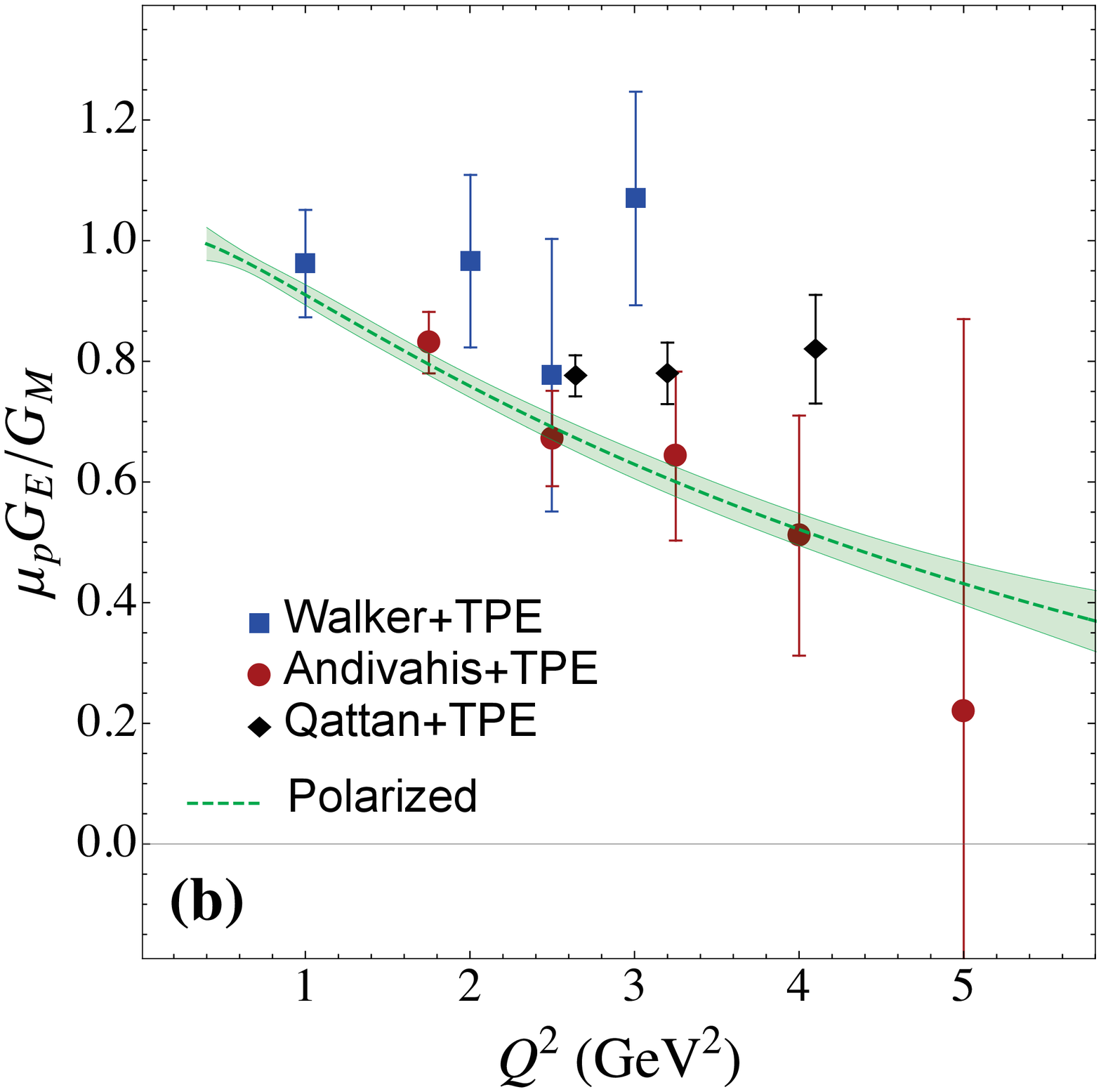}
\vspace{-3mm}
\caption{
{\bf (a)} Dependencies of ratio $R$ on $Q^2$ extracted in experiments
\cite{Andivahis1994,Walker1994,Qattan2005} using the RT.
{\bf (b)} Ratios $R$ extracted
in \cite{Blunden2020} from the reanalysis of the experiments
\cite{Andivahis1994,Walker1994,Qattan2005} using the improved RCs
from \cite{Gramolin2016} and the added TPE calculated in \cite{Blunden2020}.
Green strips correspond to the JLab polarization experiments \cite{Puckett12}.
}
\label{Blunden}
\end{figure*}

From Table \ref{Table3} and criterion (\ref{treb1}), it follows that the accuracy
of the measurements in \cite{Andivahis1994} was at a level of $1.6 \div 2.0$ \%.
This is the interval to which there also belongs the normalization uncertainty
of the Rosenbluth cross section measurement, which was 1.77 \%
(see \cite{Andivahis1994,Gramolin2016,Bernauer2014}) for all $Q^2$
in the experiment \cite{Andivahis1994}. Consequently, the measurement accuracy
for $\Delta_0$ in criterion (\ref{treb1}) should be identified with the normalization
uncertainty. At this accuracy (1.77 \%), reliability assessment criterion (\ref{treb1})
does not hold for all cells in the diagonal of Table \ref{Table3} at $Q^2\geqslant 5.0$ GeV$^2$.
Their corresponding measurements are regarded as unreliable; the values in
Table \ref{Table3} diagonal at $Q^2\geqslant 5.0$ GeV$^2$ are in bold.

\vspace{-3mm}
\begin{table*}[h!tpb] 
\caption{  
\label{Table3}
Values $\Delta_{\sigma}$ (\ref{Delta_si}) at $R=R_d$ (\ref{Rd})
 for $E_1$ (GeV) and $Q^2$ (GeV$^2$) used in the experiment \cite{Andivahis1994}
}
\tabcolsep=1.25mm
\footnotesize
\centering
\begin{tabular}{|c| c| c| c|  c| c| c| c|  c|}
\hline
$E_1 \backslash \, Q^2$
& 1.75 & 2.50 & 3.25 & 4.00 & 5.00 & 6.00 & 7.00 & 8.83 \\
\hline
9.800 & & &  & 0.097 & 0.083 & 0.067 & 0.055 & \\
\hline
5.507 & 0.197 & 0.142 & 0.107 & 0.083 & 0.060 & & & {\bf 0.006} \\
\hline
4.507 & & &  & 0.073 & 0.046 & & {\bf 0.009} & \\
\hline
3.956 & & 0.129 &  0.091 & 0.063 &  0.034 &  {\bf 0.012} & & \\
\hline
3.400 & & 0.136 & 0.085 & 0.047 & {\bf 0.015} & & & \\
\hline
2.837 & & 0.102 & 0.056 &  {0.021} & & & & \\
\hline
2.407 & 0.154 & 0.080 &  {0.028} & & & & & \\
\hline
1.968  & & 0.039 & & & & & & \\
\hline
1.511 & 0.061 & & & & & & & \\
\hline
\end{tabular}
\end{table*}

The cell at $Q^2=8.83$ GeV$^2$ and $E_1=5.507$ GeV in Table \ref{Table3} corresponds
to $\Delta_{\sigma}=0.006$, which requires measurement accuracy at a level
of $0.3 \div 0.5$ \%. However, this level of accuracy was achieved only in experiment
\cite{Bernauer2010-1} in the region where $Q^2<1$ GeV$^2$. Note that at $Q^2=8.83$ GeV$^2$
the RT-based measurement procedure in \cite{Andivahis1994} is violated, since in these
experiments the measurements for each $Q^2$ should be performed at least at two,
or better at three, electron-beam energies \cite{Bernauer2010}. A similar conclusion
about the unreliability of the measurements in \cite{Andivahis1994} at $Q^2\geqslant 5.0$
GeV$^2$ was drawn in \cite{Egle_2016}.

{\bf Reliability analysis of measurements in the experiment \cite{Qattan2005}}.
Figure \ref{Blunden}(b) also shows the results of the reanalysis of the experimental
measurements \cite{Qattan2005} with the added TPE contribution (Qattan + TPE) presented
as black diamonds. They are systematically higher than the green strip corresponding
to the results of the polarization measurements in \cite{Puckett12}. Calculations
of relative contribution $\Delta_{\sigma}$ (\ref{Delta_si}) in the kinematics of the
experiment \cite{Qattan2005} are given in Table \ref{Qattan}. Since the normalization
error in \cite{Qattan2005} was 1.7 \% \cite{Bernauer2014}, there is only one cell
in the diagonal in Table \ref{Qattan} with $E_1=2.842$ and $Q^2=4.10$ GeV$^2$
(with bold type) for which reliability assessment criterion (\ref{treb1}) does not hold.
The remaining discrepancies in \cite{Blunden2020} between Qattan + TPE and
\cite{Puckett12} are most probably caused by the underestimation of the normalization
uncertainty in \cite{Qattan2005}. The values in Table \ref{Qattan} make it possible
to conclude that it was not 1.7 but 2.0 \%. Note that, with approximate criterion
(\ref{treb3}), all measurements in \cite{Qattan2005} are classified as reliable \cite{JETPL2021}.

\vspace{-5mm}
\begin{table}[h!]
\caption{  
Values $\Delta_{\sigma}$ (\ref{Delta_si}) at $R=R_d$ (\ref{Rd}) for $E_1$ (GeV) and
$Q^2$ (GeV$^2$) used in the experiment \cite{Qattan2005}
}
\label{Qattan}
\centering
\footnotesize
\tabcolsep=1.25mm
\begin{tabular}
{| c |  c | c |  c | }
\hline
$E_1$ $\backslash \, Q^2$  & 2.64 & 3.20 & 4.10  \\
\hline
4.702 & 0.129 & 0.103 & 0.072 \\
\hline
3.772 & 0.118  & 0.090 & 0.055  \\
\hline
2.842 & 0.093 & 0.059 & {\bf 0.017}  \\
\hline
2.262 & 0.057 & {0.018}  &     \\
\hline
1.912 &  {0.020} &  &       \\
\hline
\end{tabular}
\end{table}

{\bf Reliability analysis of measurements in the experiment \cite{Walker1994}}.
Results of the calculations of relative contribution $\Delta_{\sigma}$ (\ref{Delta_si})
for all $E_1$ and $Q^2$ used in the kinematics of experiment \cite{Walker1994}
are presented in Table \ref{Walker}. Note that measurements \cite{Walker1994}
at each electron-beam energy $E_1$ were carried out in the region of small $Q^2$,
located considerably far from $Q^2_{max}$. That is why the values in Table \ref{Walker}
are not small and satisfy the inequality $\Delta_{\sigma} \geqslant 0.086$.
Since the normalization uncertainty of the measurement \cite{Walker1994}
was 1.9 \% \cite{Walker1994,Gramolin2016}, reliability assessment criterion (\ref{treb1})
holds for all values in Table \ref{Walker}. The remaining discrepancies
can be due to the fact that either the reanalysis of the experiment \cite{Walker1994}
in \cite{Blunden2020} was not quite correct, which is hardly probable, or the normalization
uncertainty in \cite{Walker1994} is underestimated by about an order of magnitude
(see Table \ref{Walker}).

\begin{table}[h!]
\caption{  
Values $\Delta_{\sigma}$ (\ref{Delta_si}) at $R=R_d$ (\ref{Rd}) for $E_1$ (GeV) and
$Q^2$ (GeV$^2$) used in the experiment \cite{Walker1994}
}
\vspace{1mm}
\label{Walker}
\centering
\tabcolsep=1.8mm
\footnotesize
\begin{tabular}
{| c |  c | c |  c | c|}
\hline
$E_1$ $\backslash Q^2$  & 1.000 & 2.000 & 2.500 & 3.000  \\
\hline
8.250 & & & 0.149 & 0.125 \\
\hline
7.500 & & 0.180  &  &   \\
\hline
7.000 & & 0.179  & 0.147 & 0.124  \\
\hline
6.250 & & 0.177 & & 0.121  \\
\hline
5.500 & & 0.175  &  &   \\
\hline
5.500 & & 0.175  &  &   \\
\hline
4.250 & & & 0.132 &        \\
\hline
4.008 & & & 0.130 &  0.103      \\
\hline
3.250 & 0.296 & 0.155 & 0.116 & 0.086        \\
\hline
2.800 & & 0.143 & 0.101 & \\
\hline
2.400 & 0.282 & 0.126 &  &   \\
\hline
1.594 & 0.238 &  &  &   \\
\hline
\end{tabular}
\end{table}

{ \bf Reliability analysis of measurements in the experiment \cite{Christy2021}}.
Results of the calculation of relative contribution $\Delta_{\sigma}$ (\ref{Delta_si})
for all $E_1$ and $Q^2$ and used in the kinematics of the experiment \cite{Christy2021}
are presented in Table \ref{Gramolin}. Measurements in \cite{Christy2021} were performed
using the left and right high-resolution spectrometers, LHRS and RHRS. In
Table \ref{Gramolin}, the values corresponding to the RHRS measurements
are marked with an asterisk. The normalization uncertainty of the Rosenbluth
cross section measurements by the LHRS and RHRS was 1.6 and 2.0 \%
respectively \cite{Christy2021}. The values in bold in Table \ref{Gramolin} are
those for which reliability assessment criterion (\ref{treb1}) does not hold. Almost
all of them, except for one, are related to the RHRS measurements and marked with
an asterisk. The only unreliable LHRS measurement corresponds to the cell with
the maximum values $E_1=10.587$ GeV and $Q^2=15.76$ GeV$^2$, where $\Delta_{\sigma}=0.009$.
For $E_1=10.587$ GeV, there are two rows in Table \ref{Gramolin}. The top row presents
the single real, but unreliable, measurement, for which $\Delta_{\sigma}=0.009$, and
the bottom row presents the missed opportunities to perform reliable measurements,
including those at $Q^2>2.0$ GeV$^2$. Thus, the kinematics used in \cite{Christy2021} is
not a good choice, and experiment \cite{Christy2021} can hardly be considered
a precision one, since about 40 \% of its measurements do not meet the reliability
assessment criterion.

\begin{table*}[h!tpb] 
\caption{   
Values $\Delta_{\sigma}$ (\ref{Delta_si}) at $R=R_d$ (\ref{Rd}) for $E_1$ $(GeV)$
and $Q^2$ (GeV$^2$) at which measurements were performed in the experiment \cite{Christy2021}.
Values with (without) an asterisk are for the RHRS (LHRS) measurements. The respective
normalization uncertainties are 2.0 and 1.6 \% }
\vspace{1mm}
\label{Gramolin}
\centering
\tabcolsep=1.5mm
\footnotesize
\begin{tabular}{|c|c| c| c|  c| c| c| c|  c|c |c| c| c| c|}
\hline
$E_1 \backslash Q^2$ &
1.577 & 1.858 & 4.543 & 5.947 & 6.993 & 7.992 & 9.002 & 9.053 & 9.807 & 11.19 & 12.07
& 12.57 & 15.76 \\
\hline
10.587 & &  &  &  &  & & & & & & & & {\bf 0.009}  \\
\hline
10.587 &0.221 & {0.193} & {0.086} & {0.064} & {0.053} & {0.045} & {0.038} & {0.038} & {0.033}
& {0.026} & {0.022} & {0.020} &  \\
\hline
8.518 &  &  &  &  &  &  &  0.031 &  & 0.026
& *{\bf 0.018}  & *{\bf 0.013}  & *{\bf 0.011}  &  \\
\hline
6.427 & &  & 0.076 & 0.051 & 0.037 & 0.026 &  & {\bf *0.016}  & & & & &  \\
\hline
2.222 &0.168 & *0.130 &  &  &  & & & & & & & & \\
\hline
\end{tabular}
\end{table*}

\section{Polarization transfer from the initial to the final proton in the elastic
$e \vec p \to e \vec p$ process }

\begin{figure*} [h!tpb]
\centering
\includegraphics[width=0.45\textwidth]{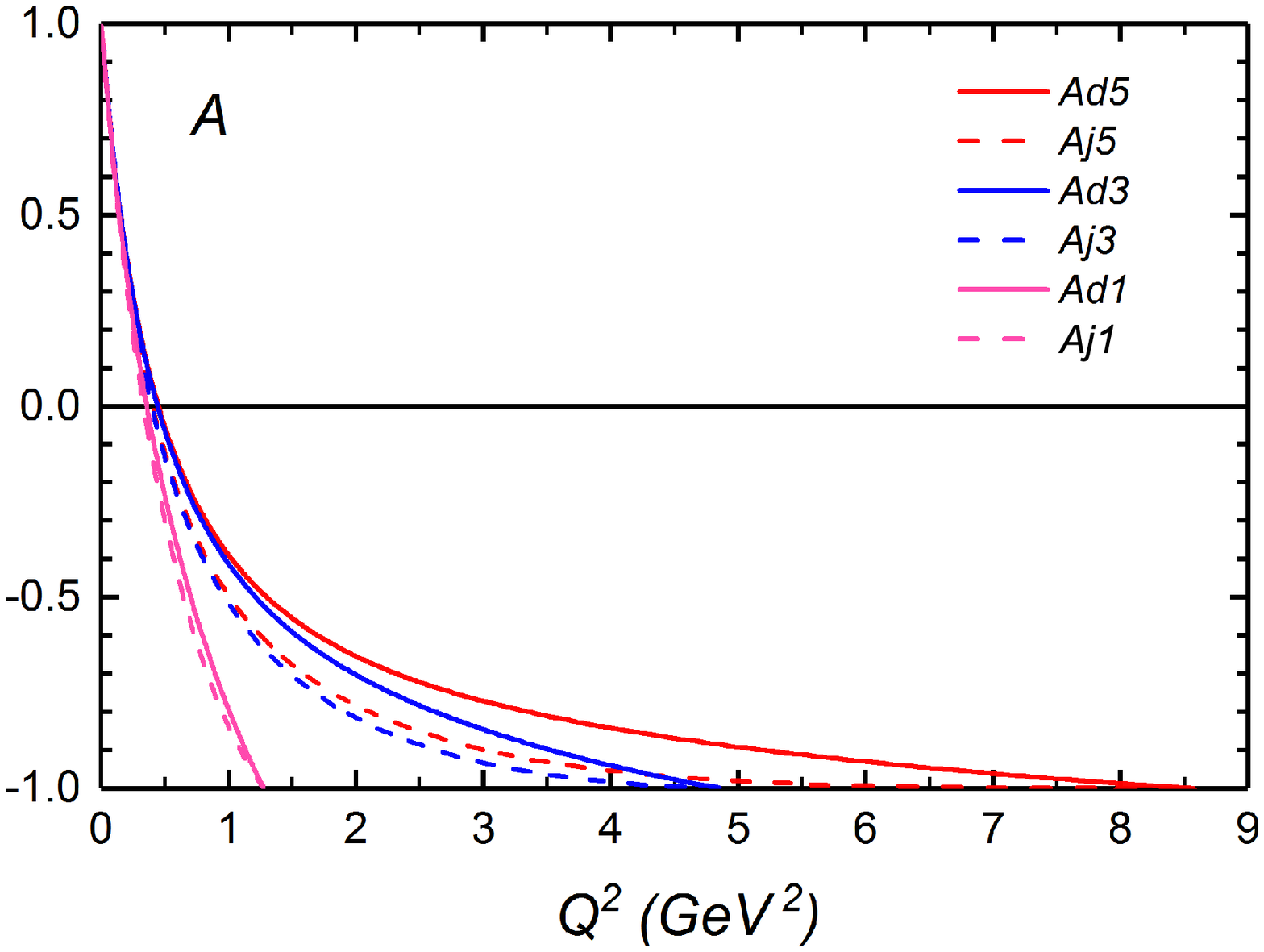}
\includegraphics[width=0.45\textwidth]{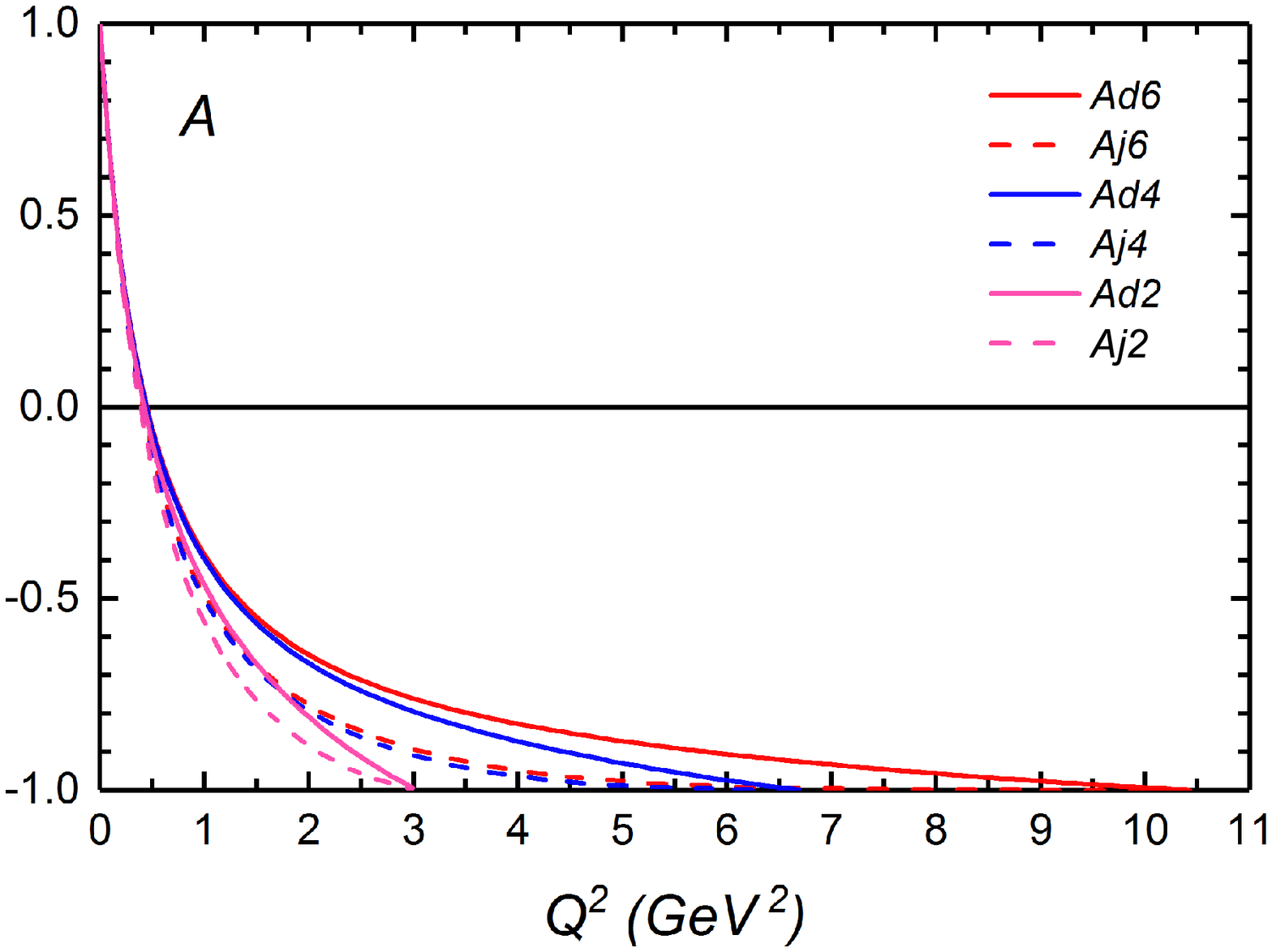}
\vspace{-3mm}
\caption{
Dependence of polarization asymmetry $A$ (\ref{Asim}) on $Q^2$ (GeV$^2$) for electron-beam
energies $E_1=1, 2, ..., 6$ GeV. Lines $Ad1$, $Ad2$, ..., $Ad6$ and $Aj1$, $Aj2$, ..., $Aj6$
correspond to ratios $R=R_d$ (\ref{Rd}) and $R=R_j$ (\ref{Rdj}), respectively.}
\label{Asimfig12}
\end{figure*}

The method proposed in \cite{JETPL18} for measuring squares of the SFFs in processes
without and with proton spin flip requires a completely polarized proton target, which seems
to be a matter for the quite distant future. As was noted above, in a wider sense it can be
considered as a method based on polarization transfer from the initial to the final proton.
The polarization transferred to the proton when the initial proton is completely polarized
($P_t=1$ ) is determined by polarization asymmetry $A$ (\ref{Asim}). To clarify generalities,
the $Q^2$ dependence of polarization asymmetry $A$ (\ref{Asim}) was numerically calculated
for the electron-beam energies $E_1=1, 2, ..., 6$ GeV. The results are shown
in Fig. \ref{Asimfig12} for $R=R_d$ (\ref{Rd}) (lines $Ad1$, $Ad2$, …, $Ad6$)
and $R = R_j$ (\ref{Rdj}) (lines $Aj1$, $Aj2$, …, $Aj6$).

It is evident from the plots in Fig. \ref{Asimfig12} that, at $P_t=1$, polarization asymmetry
$A$ (\ref{Asim}) changes, as it should, from $A=+1$ to $A=-1$, passing through $0$ at $Q^2=Q_0^2$.
At $Q^2> Q_0^2$, spin-flip cross section $\sigma^{\downarrow\uparrow}$ is larger than the non-spin-flip
cross section $\sigma^{\uparrow\uparrow}$, with their ratio being $R_{\sigma}<1$. As a result,
the helicity carried away by the recoil proton becomes negative. It reaches its maximum in absolute
value $|A|=1$ upon backward (180°) electron scattering. Note also that,
at low electron-beam energies (e.g., at $E_1=1$ GeV), the difference in the behavior
of asymmetry $A$ (\ref{Asim}) at $R=R_d$ and $R=R_j$ is insignificant. At $E_1>1$ GeV
and $Q^2>1$ GeV$^2$, the difference grows appreciable and the inequality of absolute
values $|A_j|>|A_d|$ holds.

\subsection
{Proposed experiment on measuring the SFF ratio in the $e \vec p \to e \vec p$ process}

In the general case, where the proton target is partially polarized ($P_t <1$), the degree
of longitudinal polarization transferred to the proton is defined by formula (\ref{Asim_pt}).
At present, an experiment on its measurement appears to be quite feasible, since the target
with a high degree of polarization $P_t=(70 \pm 5)$ \% has in principle been developed
and was already used in \cite{Liyanage2020}. It is for this reason that the proposed experiment
should preferably be conducted at the facility used by the SANE collaboration \cite{Liyanage2020}
at the same $P_t=0.70$, electron-beam energies $E_1=4.725$ and 5.895 GeV, and
squares of momenta transferred to the proton $Q^2 = 2.06$ and 5.66 GeV$^2$. The difference
between the proposed experiment and \cite{Liyanage2020} is that the electron-beam should
be unpolarized and the detected recoil proton should move strictly along the spin
quantization axis of the proton target. Degrees of longitudinal and transverse polarization
of the final proton were measured in \cite{Jones00,Gay01,Gay02,Pun05,Puckett10,Puckett12}.
In the proposed experiment, it is necessary to measure only the degree of longitudinal
polarization of the recoil proton, which is an advantage when compared to the method
\cite{Rekalo74} used in the JLab experiments.

\begin{figure}[h!]
\centering
\includegraphics[scale=0.35]{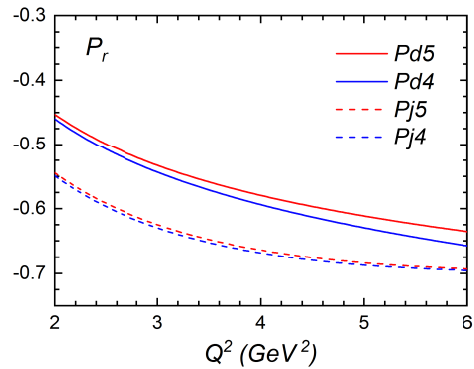}
\vspace{-6mm}
\caption{
Dependence of the degree of longitudinal polarization
of the recoil proton $P_r$ (\ref{Asim_pt}) on the square of the
momentum transferred to the proton $Q^2$ (GeV$^2$) for $E_1$ and $P_t$
used in \cite{Liyanage2020}. Lines $Pd5$, $Pd4$ (solid) and $Pj5$, $Pj4$
(dashed) correspond to ratios $R=R_d$ (\ref{Rd}) and $R=R_j$ (\ref{Rdj}).
}
\label{exp}
\end{figure}

The results of calculating the $Q^2$ dependence of polarization transferred to proton
$P_r$ (\ref{Asim_pt}) in the kinematics of the experiment \cite{Liyanage2020} are shown
in Fig. \ref{exp}, where lines $Pd5$, $Pd4$ (solid) and $Pj5$, $Pj4$ (dashed)
are constructed for relations $R=R_d$ (\ref{Rd}) and $R=R_j$ (\ref{Rdj}). Lines $Pd5$, $Pj5$,
correspond to the electron-beam
energy $E_1=5.895$ GeV, and lines $Pd4$, $Pj4$, correspond to $E_1=4.725$ GeV.
The degree of proton target polarization is $P_t=0.70$ for all lines in Fig. \ref{exp}.

\vspace{-5mm}

It follows from Fig. \ref{exp} that polarization transferred to the recoil proton depends
appreciably on the form of the dependence of ratio $R$ on $Q^2$. In the case of SFFs
scaling violation, i.e., when $R=R_j$, it noticeably increases in absolute value
when compared to the case where $R=R_d$; i.e., inequalities $|Pj5|>|Pd5|$ and  $|Pj4|>|Pd4|$
hold for all $Q^2$. A quantitative estimation of this difference is given in Table \ref{Table7},
which presents degrees of longitudinal polarization of the final proton $Pj5$, $Pd5$, $Pj4$,
and $Pd4$ and their relative difference $\Delta_{dj5}$ and $\Delta_{dj4}$
(in percent) at two electron-beam energies of  5.895 and 4.725 GeV and two $Q^2$
of 2.06 and 5.66 GeV$^2$, where $\Delta_{dj5}=(Pj5-Pd5)/Pj5$ and $\Delta_{dj4}=(Pj4 - Pd4)/Pj4$.

\begin{table}[h!tpb] 
\centering
\caption{  
Degree of longitudinal polarization of the recoil
proton $P_r$ (\ref{Asim_pt}) at electron-beam energies $E_1=5.895$ and
4.725 GeV and $Q^2=2.06$ and 5.66 GeV$^2$
}
\vspace{1mm}
\label{Table7}
\tabcolsep=1.20mm
\footnotesize
\begin{tabular}
{| c | c | c | c | c | c | c | c |}
\hline
$Q^2$ (\rm{GeV}$^2$) & $Pd5$ & $Pj5$ & $Pd4$ & $Pj4$ & $\Delta_{dj5}$, \%  & $\Delta_{dj4}$, \%  \\
\hline
2.06 & -- 0.46 & -- 0.55 & -- 0.47 & -- 0.56 & 16.6 & 16.1 \\
\hline
5.66 & -- 0.63 & -- 0.69 & -- 0.65 & -- 0.69 & 9.1 & 6.4   \\
\hline
\end{tabular}
\end{table}

It follows from Table \ref{Table7} that at $Q^2=2.06$ GeV$^2$ the relative difference
between $Pj5$ and $Pd5$ is 16.6 \%, and between $Pj4$ and $Pd4$ it is approximately the same,
16.1 \%. At $Q^2 = 5.66$ GeV$^2$ this difference decreases to 9.1 and 6.4 \% respectively.

Note that, after the measurement of the degree of longitudinal polarization of the recoil
proton, $R$  is extracted using relation (\ref{Rp}).

\vspace{-3mm}
\section*{CONCLUSIONS}

In this work, a reliability assessment criterion for measurements of the
ratio $R$ using the RT is proposed, according to which the relative contribution
of the $G^{\,2}_E$ involving term $\sigma^{\uparrow\uparrow}$ to the Rosenbluth
cross section should be larger than the normalization uncertainty of the measurement
of this cross section. Based on the results of the reanalysis \cite{Blunden2020},
a reliability analysis was performed using this criterion for measurements in the
known experiment \cite{Andivahis1994,Qattan2005,Walker1994} and for the recent experiment
\cite{Christy2021} at JLab’s CEBAF accelerator upgraded to 12 GeV. It follows
from the analysis that, first, the measurements \cite{Andivahis1994}
at  $Q^2>5$ GeV$^2$ are unreliable.
Second, the remaining discrepancies between the measurements
\cite{Qattan2005} (with added TPE contribution) and the
results of the polarization experiments \cite{Puckett12} observed in
\cite{Blunden2020} can result from the normalization uncertainty of the
Rosenbluth cross section measurement in \cite{Qattan2005} being
underestimated, being not 1.7 \% but 2.0 \%.
Third, the remaining discrepancies found in \cite{Blunden2020} for the measurements
in \cite{Walker1994} can be due to the fact that either the
reanalysis of the experiment \cite{Walker1994} in \cite{Blunden2020} was not quite
correct, which is hardly probable, or the uncertainties of the
Rosenbluth cross section measurements in \cite{Walker1994} are
underestimated by about an order of magnitude.
Fourth, the kinematics of the experiment \cite{Christy2021} is not
quite a good choice, because about 40 \% of the measurements
performed in it are not reliable.

Based on the results of the JLab polarization experiments on measurement
of the ratio $R$ in the $\vec e  p \to e \vec p$ process, a numerical analysis
was performed for the dependence of the ratio of the cross sections without
and with proton spin flip on the square of the momentum transferred to the proton
and for the polarization asymmetry in the process when the initial (at rest) and
final protons are completely polarized and have a common spin quantization
axis coinciding with the direction of motion of the detected recoil proton.
When the initial proton is partially polarized, the longitudinal polarization
transferred to the proton is calculated in the kinematics used by the SANE collaboration
\cite{Liyanage2020} in the experiments on measuring double
spin asymmetry in the $\vec e \vec p \to e p$ process. The polarization
transferred to the proton is found to be noticeably
sensitive to the form of the ratio $R$ dependence on $Q^2$,
which can be useful for a new independent experiment
on its measurement in the $e \vec p \to e \vec p$ process.

I thank R. Lednicky for interest in this work and helpful
discussions of the results.

\end{document}